\documentclass[twocolumn,superscriptaddress,amsmath, amssymb, amsfonts,preprintnumbers,aps,prd,longbibliography,nofootinbib, 10pt]{revtex4-1}

\usepackage{scalerel}
\usepackage[dvipsnames]{xcolor}
\usepackage{amsmath,amsfonts,amssymb}
\usepackage[small,bf,hang]{caption}
\usepackage{slashed}
\usepackage{latexsym,epsfig}
\usepackage{dsfont}
\usepackage{arydshln}
\usepackage{extarrows}
\usepackage{hyperref}

\usepackage{tikz}
\usepackage{tikz-cd}
\usetikzlibrary{decorations.pathmorphing}
\usepackage{mathtools}
\tikzset{snake it/.style={decorate, decoration=snake}}
\usetikzlibrary {shapes.geometric} 
\usetikzlibrary{shapes.misc}
\usetikzlibrary{decorations.markings}
\pgfdeclarelayer{edgelayer}
\pgfdeclarelayer{nodelayer}
\pgfsetlayers{edgelayer,nodelayer,main}

\tikzstyle{imu}=[fill={rgb,255: red,22; green,188; blue,119}, draw={rgb,255: red,22; green,188; blue,119}, shape=circle, scale=0.5]
\tikzstyle{theta 2}=[fill=cyan, draw=cyan, shape=circle,scale=0.5]
\tikzstyle{filter}=[draw,shape=cross out,rotate=42,scale=0.7]
\tikzstyle{m2}=[fill={rgb,255: red,128; green,0; blue,128}, draw=none, shape=circle,scale=0.5]
\tikzstyle{f2}=[fill=cyan, draw=cyan, shape=rectangle,scale=0.7, scale=0.8]
\tikzstyle{mu 2}=[fill={rgb,255: red,184; green,132; blue,183}, draw={rgb,255: red,184; green,132; blue,183}, shape=circle,scale=0.5]
\tikzstyle{partial mu}=[fill={rgb,255: red,128; green,128; blue,128}, draw={rgb,255: red,128; green,128; blue,128}, shape=circle,scale=0.5]
\tikzstyle{Q}=[fill=none, draw=black, shape=circle]
\tikzstyle{root filter}=[draw,shape=cross out]
\tikzstyle{b2}=[fill={rgb,255: red,222; green,51; blue,45}, draw={rgb,255: red,222; green,51; blue,45}, shape=rectangle,scale=0.8]
\tikzstyle{s2}=[fill={rgb,255: red,255; green,191; blue,191}, shape=rectangle, scale=0.8]
\tikzstyle{dp}=[draw,rectangle, rotate=45, scale=0.8]
\tikzstyle{arrow}=[->]
\tikzstyle{root filter}=[draw,shape=cross out]
\tikzstyle{off shell}=[fill=none, draw=black, shape=circle,scale=0.5]

\tikzstyle{M}=[-, fill={rgb,255: red,192; green,135; blue,241}, draw={rgb,255: red,192; green,135; blue,241}, line width=2pt]
\tikzstyle{c}=[-, fill={rgb,255: red,191; green,0; blue,64}, draw={rgb,255: red,191; green,0; blue,64},line width=2pt]
\tikzstyle{cut}=[dashed, color=WildStrawberry]

\def\moth{\mathsurround=0pt}
\newdimen\zo \zo=0pt

\def\tick{\leaders\hrule height 0.5ex depth 0pt \hskip 0.5pt}
\def\upboxfill{$\moth \setbox\zo\hbox{\tick}%
  \hskip 3pt\hbox to 0pt{$\tick$\hss}\hrulefill \hbox to 7.5pt{$\tick$\hss}$}

\def\dtick{\leaders\hrule height .34pt depth 0.5ex \hskip 0.5pt}
\def\downboxfill{$\moth \setbox\zo\hbox{\dtick}%
  \hskip 2pt\hbox to 0pt{$\dtick$\hss}\hrulefill \hbox to 2pt{$\dtick$\hss}$}

\def\cA{{\cal A}}
\def\cE{{\cal E}}
\def\cM{{\cal M}}

\def\cK{{\cal K}}

\def\cI{{\cal I}}

\def\del{\partial}

\def\B{\square}

\begin{document}

\preprint{
HU-EP-26/01-RTG
}
\title{Color-kinematics duality from an algebra of superforms  }

\author{Roberto Bonezzi} 
\email{roberto.bonezzi@physik.hu-berlin.de}
\affiliation{%
Institut f\"ur Physik und IRIS Adlershof, Humboldt-Universit\"at zu Berlin,
Zum Gro{\ss}en Windkanal 2, 12489 Berlin, Germany
}

\author{Christoph Chiaffrino} 
\email{chiaffrc@hu-berlin.de}
\affiliation{%
Institut f\"ur Physik und IRIS Adlershof, Humboldt-Universit\"at zu Berlin,
Zum Gro{\ss}en Windkanal 2, 12489 Berlin, Germany
}

\author{Olaf Hohm} 
\email{ohohm@physik.hu-berlin.de}
\affiliation{%
Institut f\"ur Physik und IRIS Adlershof, Humboldt-Universit\"at zu Berlin,
Zum Gro{\ss}en Windkanal 2, 12489 Berlin, Germany
}

\author{ Maria Foteini Kallimani} 
\email{kallimari@physik.hu-berlin.de}
\affiliation{%
Institut f\"ur Physik und IRIS Adlershof, Humboldt-Universit\"at zu Berlin,
Zum Gro{\ss}en Windkanal 2, 12489 Berlin, Germany
}

\begin{abstract}

Color-kinematics duality states that  the kinematic numerators of the cubic tree-level Yang-Mills scattering amplitudes obey the same symmetry properties that the color factors obey due to 
the Jacobi identity. 
We present a novel strategy for  deriving this duality, based on the differential forms on a superspace. 
This space of superforms 
carries a generalization of a Batalin-Vilkovisky (BV) algebra  (BV$^{\square}$ algebra). We show 
that the homotopy algebra of color-stripped Yang-Mills theory is  obtained as a quotient of this space 
in which a subspace, which is an ideal `up to homotopy', is modded out. 
This algebra is a subsector of 
a BV$_{\infty}^{\square}$ algebra. 
Deriving the latter would provide a first-principle proof of color-kinematics duality from field theory.

\end{abstract}

\maketitle

\section{Introduction}

The known  fundamental forces in nature are governed by Yang-Mills theory in  particle physics on the one hand 
and by general relativity in  gravitational physics on the other.  Both theories share broad qualitative features 
such as gauge invariance, but their detailed Lagrangians  appear radically different. Yang-Mills theory is governed 
by a Lagrangian with a cubic vertex, which is linear in derivatives, and a quartic vertex, which is independent of derivatives. 
This leads to a renormalizable quantum field theory (QFT). In contrast, general relativity expanded about Minkowski space 
yields interaction vertices of arbitrary order, each featuring two derivatives. This yields a non-renormalizable QFT. 
 
Nevertheless, it was established by Bern, Carrasco and Johansson (BCJ) \cite{Bern:2008qj,Bern:2010ue} that there is a deep relation between the
tree-level scattering amplitudes of a certain gravity theory and that of Yang-Mills theory. To state this relationship  one writes 
a tree-level  $n$-point scattering amplitude of Yang-Mills theory as the Feynman diagram  expansion, 
but only using  \textit{cubic} diagrams $\Gamma$: 
\begin{equation}\label{Amplitude}
\cA_n=g^{n-2}\sum_{\Gamma}\frac{c_{\Gamma}\,n_\Gamma}{D_\Gamma}\;. 
\end{equation}
This can be achieved, for instance,  by replacing  the field strength $F_{\mu\nu}$ by  an independent auxiliary field
or  by suitably ``blowing up'' the quartic vertices, so that all Feynman rules become cubic. 
Moreover, $c_\Gamma$ is the color factor associated with a given diagram $\Gamma$, consisting 
of $n-2$ structure constants $f^a{}_{bc}$, $n_\Gamma$ is the kinematic numerator built 
from the $n$ polarization vectors $\epsilon_i^\mu$ and momenta $k_i^\mu$, 
and $D_{\Gamma}$ encodes  the inverse propagators. 
The BCJ or color-kinematics (CK) 
duality states that whenever $c_{\Gamma_1}+c_{\Gamma_2}+c_{\Gamma_3}=0$ holds as a consequence 
of the Jacobi identity of the color gauge group, then the corresponding numerators obey 
$n_{\Gamma_1}+n_{\Gamma_2}+n_{\Gamma_3}=0$. Once the numerators $n_\Gamma$ have this CK duality obeying 
property one may replace in (\ref{Amplitude}) the color factors $c_\Gamma$ by a second copy $\tilde{n}_\Gamma$ of 
the numerators. This BCJ double copy yields the scattering amplitudes of ${\cal N}=0$ supergravity (Einstein gravity coupled to 
a 2-form B-field and a scalar dilaton). 

The CK duality and the associated double copy nature of gravity express  a deep relationship between Yang-Mills theory and gravity \cite{Bern:2010yg,Bern:2019prr,Bern:2022wqg,Adamo:2022dcm}. 
While there are indirect proofs from string theory of CK duality at tree-level \cite{Bjerrum-Bohr:2010pnr,Mafra:2011kj}, it is desirable to obtain a first-principle 
derivation directly from field theory, in particular to obtain an effective algorithm to compute the 
kinematic numerators in a manifestly local and CK duality compatible  fashion.  This latter feature in particular  will be important 
 in order to obtain full control of CK duality and the double copy at 
 loop level. 

A natural idea to prove CK duality from first principles  is to identify a hidden `kinematic Lie algebra', in which case  $n_{\Gamma_1}+n_{\Gamma_2}+n_{\Gamma_3}=0$ would follow from its Jacobi identity \cite{Monteiro:2011pc,Chen:2019ywi,Chen:2021chy,Brandhuber:2021bsf}. 
Results  in recent years have shown, however,  that a more sophisticated algebraic structure 
is at the heart of CK duality \cite{Reiterer:2019dys,Ben-Shahar:2021zww,Bonezzi:2022yuh,Borsten:2022vtg,Bonezzi:2022bse,Borsten:2023reb,Bonezzi:2023pox,Ben-Shahar:2024dju}. 
To this end one uses the formulation of field theory in terms of homotopy algebras such as homotopy Lie or 
$L_{\infty}$ algebras \cite{Zwiebach:1992ie,Lada:1992wc,Zeitlin:2007vv,Zeitlin:2007vd,Zeitlin:2007ttl,Hohm:2017pnh,Jurco:2018sby,Grigoriev:2023lcc} or homotopy associative commutative or $C_{\infty}$ algebras \cite{Zeitlin:2008cc}. A $C_{\infty}$ algebra 
is defined by maps $m_k$, $k=1,2,3,\ldots$, with $k$ inputs, satisfying certain identities. 
For instance, the algebra of differential forms on any space forms a $C_{\infty}$ algebra, where 
$m_1=d$ is the de Rham differential and $m_2$ is the wedge product, while all $m_k$ for $k\geq 3$ 
are zero. Such a $C_{\infty}$ algebra is called strict. Upon taking the tensor product with a 
color Lie algebra $\frak{g}$ this gives rise, on a 3-manifold, to topological Chern-Simons theory. 
Yang-Mills theory 
is  similarly  encoded in a genuine  $C_{\infty}$ algebra with a non-trivial $m_3$. 
The action reads  
\begin{equation}\label{cinftyAction} 
\begin{split} 
 S_{\rm YM}  =&\, \frac{1}{2}\big\langle A_a, m_1(A^a)\big\rangle +\frac{1}{3!}f_{abc}  \big\langle A^a, m_2(A^b, A^c)\big\rangle\\
 &+\frac{1}{12}f_{abe} f_{cd}{}^{e}  \big\langle A^a, m_3(A^b, A^c, A^d)\big\rangle  \,,
\end{split} 
\end{equation} 
where the color factors $f_{abc}$ are  stripped off explicitly. 
In the case that the $C_{\infty}$ algebra is the strict algebra of differential forms on a 3-manifold, 
for which the quartic terms in the second line vanish, this is the Chern-Simons action. 
When instead the maps $(m_1,m_2,m_3)$ define  the color-stripped Yang-Mills  $C_{\infty}$ algebra, 
the above is the familiar Yang-Mills action. 

In this letter we show that, remarkably, the $C_{\infty}$ algebra of Yang-Mills theory can be obtained 
from the algebra of differential forms on the superspace that extends spacetime by one odd 
coordinate. To this end we identify a subspace of this  space of superforms as an `ideal' that is modded out.
More precisely, while this subspace is not an ideal of the strict algebra of differential forms, it is an 
ideal `up to homotopy', and the quotient space then inherits the  Yang-Mills $C_{\infty}$ algebra. 

It should be emphasized, however,  that this algebra is only the first layer of the full homotopy algebra
that is needed to explain CK duality \cite{Reiterer:2019dys,Bonezzi:2022bse}. 
This algebra is a generalization of a ${\rm BV}_\infty$ algebra \cite{ValetteBV} called BV$_{\infty}^{\square}$.  
In order to define this algebra, one 
uses that there is a second differential $b$, satisfying $b^2=0$, in terms of which the propagator is $\frac{b}{\square}$. While the full propagator is non-local, in our formulation $b$ is local and can be used to define 
a local Lie-type bracket $b_2$
as the failure of $b$ to act as a  derivation on $m_2$.  This  in turn gives rise to an infinite tower of higher (local) maps. The idea of deriving  the maps of the BV$_{\infty}^{\square}$ algebra from a strict algebra  was pursued  previously in \cite{Bonezzi:2024fhd,Bonezzi:2025bgv} using algebras of worldline vertex operators. What makes the new approach introduced here particularly promising is that we explicitly 
identify a strict BV$^{\square}$  algebra 
for  which we conjecture that the 
full BV$_{\infty}^{\square}$ algebra is obtained as a suitable quotient. While 
in contrast to the $C_{\infty}$ subsector we do not yet know how to construct the full BV$_{\infty}^{\square}$ algebra, we will use the opportunity to explain in sec.~IV how it encodes CK duality and comment in the conclusion section V on the ongoing program for deriving this  algebra.

\section{Algebra of superforms  } 

\begin{figure*}[ht]
   \begin{tikzpicture}
	\begin{pgfonlayer}{nodelayer}
		\node    (0) at (0, 2) {\textcolor{ForestGreen}{$(0,0)$}};
		\node    (1) at (-2, 0) {\textcolor{ForestGreen}{$(1,0)$}};
		\node    (2) at (2, 0) {\textcolor{ForestGreen}{$(0,1)$}};
		\node    (4) at (-0.25, 1.75) {};
		\node    (5) at (-1.75, 0.25) {};
		\node    (6) at (-2.25, -0.25) {};
		\node    (9) at (-4, -2) {$(2,0)$};
		\node    (10) at (-3.75, -1.75) {};
		\node    (11) at (-4.25, -2.25) {};
		\node    (12) at (-6, -4) {$(3,0)$};
		\node    (13) at (-6.25, -4.25) {};
		\node    (14) at (-5.75, -3.75) {};
		\node    (15) at (0.25, 1.75) {};
		\node    (16) at (1.75, 0.25) {};
		\node    (17) at (2.25, -0.25) {};
		\node    (18) at (0, -2) {\textcolor{ForestGreen}{$(1,1)$}};
		\node    (21) at (4, -2) {\textcolor{gray!60}{$(0,2)$}};
		\node    (22) at (3.75, -1.75) {};
		\node    (23) at (4.25, -2.25) {};
		\node    (24) at (6, -4) {\textcolor{gray!60}{{$(0,3)$}}};
		\node    (25) at (5.75, -3.75) {};
		\node    (26) at (6.25, -4.25) {};
		\node    (27) at (-2, -4) {$(2,1)$};
		\node    (28) at (-2.25, -3.75) {};
		\node    (29) at (-2, -4.25) {};
		\node    (33) at (2, -4) {\textcolor{ForestGreen}{$(1,2)$}};
		\node    (35) at (2, -4.25) {};
		\node    (36) at (-9, 2) {};
		\node    (37) at (-9, 2) {$G=0$};
		\node    (38) at (-9, 0) {$G=1$};
		\node    (39) at (-9, -2) {$G=2$};
		\node    (40) at (-9, -4) {$G=3$};
		\node    (41) at (1.75, -0.25) {};
		\node    (42) at (0.25, -1.75) {};
		\node    (43) at (3.75, -2.25) {};
		\node    (44) at (2.25, -3.75) {};
		\node    (45) at (-3.75, -2.25) {};
		\node    (46) at (-0.25, -1.75) {};
		\node    (47) at (-1.75, -0.25) {};
		\node    (48) at (-1.75, -3.75) {};
		\node    (49) at (1.75, -3.75) {};
		\node    (50) at (-0.25, -2.25) {};
		\node    (51) at (0.25, -2.25) {};
		\node    (61) at (-1.25, 1.25) {$d$};
		\node    (63) at (1.25, 1.25) {$\Delta$};
		\node    (64) at (-3.25, -0.75) {$d$};
		\node    (65) at (-5.25, -2.75) {$d$};
		\node    (66) at (3.25, -0.75) {$\Delta$};
		\node    (67) at (5.5, -3) {$\Delta$};
		\node    (68) at (-0.75, -0.75) {$\Delta$};
		\node    (69) at (-2.75, -2.75) {$\Delta$};
		\node    (70) at (1.25, -2.75) {$\Delta$};
		\node    (71) at (0.75, -0.75) {$d$};
		\node    (72) at (2.75, -2.75) {$d$};
		\node    (73) at (-1.25, -2.75) {$d$};
	\end{pgfonlayer}
	\begin{pgfonlayer}{edgelayer}
		\draw [style=arrow] (4.center) to (5.center);
		\draw [style=arrow] (6.center) to (10.center);
		\draw [style=arrow] (11.center) to (14.center);
		\draw [style=arrow] (15.center) to (16.center);
		\draw [style=arrow] (45.center) to (28.center);
		\draw [style=arrow] (47.center) to (46.center);
		\draw [style=arrow] (51.center) to (49.center);
		\draw [style=arrow] (17.center) to (22.center);
		\draw [style=arrow] (23.center) to (25.center);
		\draw [style=arrow] (43.center) to (44.center);
		\draw [style=arrow] (41.center) to (42.center);
		\draw [style=arrow] (50.center) to (48.center);
	\end{pgfonlayer}
\end{tikzpicture}
\caption{Bicomplex of superforms. The degrees $(p,q)$ in the above diagram are given by $(N,G-N)$. The spaces that descend to the Yang-Mills kinematic algebra $\cK$ are displayed in green. Spaces in gray are zero, due to $c^2=0$.}
\label{bicomplex}
\end{figure*}
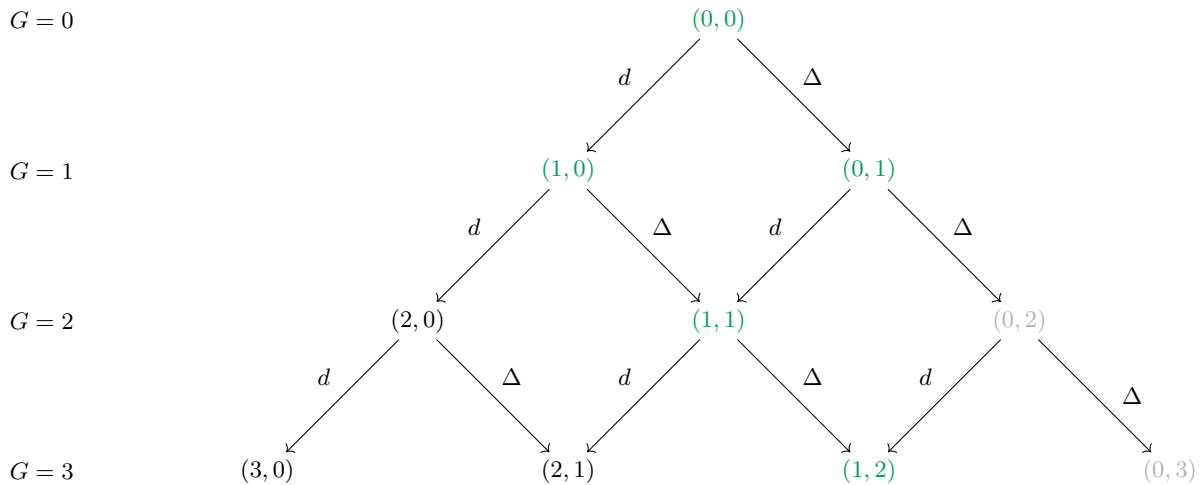

We consider the algebra  of differential forms (the de Rham complex) on the superspace $\mathbb{R}^{D|1}$ 
with coordinates  $(x^{\mu}, c)$, where $x^{\mu}$ are the bosonic spacetime coordinates and $c$  is an odd 
coordinate satisfying $c^2=0$. This type of construction has been previously studied in \cite{Tsygan}. Viewing the basis one-forms as new (odd and even) coordinates via 
  \begin{equation}\label{CmuM} 
   C^\mu\equiv dx^\mu\;, \qquad  \cM\equiv dc  \;, 
 \end{equation} 
where  $C^\mu C^\nu=-C^\nu C^\mu$,  
the de Rham complex  $\Omega^{\bullet}(\mathbb{R}^{D|1})$ becomes the space of functions or superfields 
 \begin{equation}\label{functions}
  F(x^\mu,c\,;C^\mu,\cM)\;, 
 \end{equation} 
whose dependence on $c, C^{\mu}, \cM$ is polynomial.  
 The degrees (ghost number $G$) of the coordinates are given by 
$G(x^\mu)=0$, $G(c)=+1$, $G(C^\mu)=+1$, and $G(\cM)=+2$. Note in particular 
that since ${\cal M}$ is bosonic the functions (\ref{functions}) may depend on arbitrary powers of $\cM$. 
This space is actually a bi-complex, as depicted in Figure \ref{bicomplex}: it decomposes according to two degrees, the ghost number $G$ 
and the superform degree $N$, which counts the number of $C^\mu$ and $\cM$. 

The wedge product $\wedge$ of differential superforms is encoded in the pointwise product of superfields (\ref{functions}), while 
the de Rham differential can be written as a vector field 
\begin{equation}
 d:=C^\mu\del_\mu+\cM\frac{\del}{\del c}\;, 
\end{equation}
which satisfies $d^2=0$ and acts on the product via the Leibniz rule. 
The space of superforms $(\Omega^{\bullet}, d, \wedge)$, where 
we use the shorthand $\Omega^{\bullet}\equiv \Omega^{\bullet}(\mathbb{R}^{D|1})$ in the following, 
 forms a differential graded (dg) commutative associative algebra (dgCom). 

Assuming the existence of a metric, the above space carries further structure:  
a (strict) BV$^\B$ algebra. 
To define  this recall  the adjoint of $d$, 
 \begin{equation}\label{DEEERHAM} 
d^\dagger := \frac{\del^2}{\del x_{\mu} \del C^\mu}\;, 
\end{equation}
which obeys $[d,d^\dagger]=\B$, 
where  $\B$ is the  d'Alembert operator.  $d^\dagger$ is a  second-order operator and hence does not obey the Leibniz rule w.r.t.~$\wedge$; rather, its failure defines a (graded) Lie bracket, the Schouten-Nijenhuis bracket. 
It was shown by Ben-Shahar and Johansson that this BV$^\B$ algebra $(\Omega^{\bullet}, d, d^\dagger, \wedge)$ explains CK duality for the toy model of Chern-Simons theory \cite{Ben-Shahar:2021zww}.

We now apply an isomorphism to the above algebra that brings it to an equivalent form in 
which  the Yang-Mills kinematic space is more easily identified.  
This isomorphism 
is written as $e^h$, with inverse $e^{-h}$, where
\begin{equation}\label{hDEF} 
h:=c\,d^\dagger 
\;, 
\end{equation}
which obeys $h^2=0$ and $G(h)=0, N(h)=-1$. 
The isomorphism transports the de Rham differential   to 
\begin{equation}
Q:=e^hd\,e^{-h}=d+\Delta \;, 
\end{equation}
where $\Delta$ is a second differential given by 
\begin{equation}
\Delta =c\,\B-\cM\,d^{\dagger} \;,   
\end{equation}
which obeys $\Delta^2=0$ and (anti)commutes with $d$. 
The wedge product is transported to 
\begin{equation}
\begin{split}
\mu&:=e^h\circ\wedge\circ(e^{-h}\otimes e^{-h})\;, 
\end{split}
\end{equation}
which explicitly  is given by 
\begin{equation}\label{mu explicit}
\mu(F_1,F_2)=F_1\cdot F_2+c \left[\frac{\del F_1}{\del C^\mu}\cdot \del^\mu F_2
\pm (1\leftrightarrow 2) 
\right] . 
\end{equation}
Thus, the transported product is the wedge product  shifted by a BV antibracket, which coincides with 
the above mentioned Schouten-Nijenhuis bracket. 

Since above we applied an isomorphism, the resulting algebra $(\Omega^{\bullet},Q, \mu)$ is still 
a strict dgCom algebra. Moreover, it also carries a strict BV$^\B$ algebra, which arises as follows. 
We define the vector field 
\begin{equation}\label{cVectorfield} 
b:=\frac{\del}{\del c}\;,   
\end{equation}
which is manifestly  first-order with respect to the wedge product and commutes with $d$. However, 
with respect to the transported $\mu$, $b$ turns out to be \textit{second-order} and to have the 
following (anti)commutator with $Q$: 
 \begin{equation}\label{QbBox} 
  [Q,b]=\B\;. 
 \end{equation} 
These properties can be inferred from the identity 
\begin{equation}
\frac{\del}{\del c} = e^h\left(\frac{\del}{\del c}+ d^\dagger \right)e^{-h}  
\end{equation}
and the fact that  
$d^\dagger$ has the same properties. 
Thus, the dgCom algebra $(Q,\mu)$ is here augmented by a second differential $b$ that is second-order with 
respect to $\mu$ and obeys (\ref{QbBox}), i.e., $(\Omega^{\bullet},Q, b, \mu)$ 
 is a (strict) BV$^\B$ algebra.

\section{Ideal and quotient  } 

We now decompose the de Rham complex of superforms into a subspace that will be related to the kinematic 
space of Yang-Mills theory and its complement. In terms of  the super-form degree $N$, counting the number 
of $C^{\mu}$ and ${\cal M}$,  the de Rham complex $\Omega^{\bullet}$ 
decomposes as 
\begin{equation}
\bigoplus_{n=0}^\infty\Omega^{n}
=\Omega^{0}\oplus \Omega^{1}\oplus\cI\;,
\quad
 \cI = \bigoplus_{n=2}^\infty\Omega^{n}\;, 
\end{equation}
where $\cI$ 
defines the subspace that will become an ideal. 
More precisely, this space is a sub-chain complex of $(\Omega^{\bullet},Q)$, because 
$Q$ maps this subspace  to itself: 
\begin{equation}
\forall  \chi\in\cI:\qquad 
Q\chi\in\cI  \;.    
\end{equation}
This in turn allows us to define the Yang-Mills kinematic space  as the quotient space
\begin{equation}\label{Kquotient} 
\cK \ := \ {\Omega^{\bullet}}\mathbin{/}{\cI}   
\ \simeq \  \Omega^{0}\oplus \Omega^{1}\;. 
\end{equation}
Its elements are the  equivalence classes $[F]=[F+\chi]$ for any $\chi\in\cI$. One can then define the differential on $\cK$ via
\begin{equation}
Q[F]:=[QF]\;,    
\end{equation}
where we use the same symbol $Q$. 
$(\cK,Q)$ is the  chain complex of color-stripped Yang-Mills theory as in \cite{Bonezzi:2022yuh,Bonezzi:2022bse}. More explicitly, the single space at $G=0$ consists of color-stripped gauge parameters $\lambda(x)$, while the $(1,0)$ and $(0,1)$ spaces at $G=1$ contain the fields $C^\mu A_\mu(x)$ and $c\,\varphi(x)$, respectively, where $\varphi$ is an auxiliary scalar. The rest of $\cK$ consists of the super one-forms in the spaces $(1,1)$ and $(1,2)$, where $(1,1)$ contains the vector and scalar equations as $c\,C^\mu E_\mu(x)+\cM E(x)$, while the Noether identities $c\,\cM N(x)$ belong to the space $(1,2)$.  

The quotient space (\ref{Kquotient}) does not inherit  a 
strict BV$^\B$ algebra from $(\Omega^{\bullet},Q, b, \mu)$, 
because ${\cI}$ is not an ideal. 
For instance, the transported wedge product $\mu$ of a general element with an element in $\cI$
is generally not in $\cI$. However, for our purposes it suffices that 
$\cI$ is an ideal `up to homotopy' so that the quotient space inherits a homotopy version of a BV$^\B$ algebra: 
a BV$_{\infty}^\B$ algebra. By this we will mean that one can apply a suitable morphism of BV$_{\infty}^\B$ algebras to 
$(\Omega^{\bullet},Q, b, \mu)$, after which $\cI$ becomes an ideal in the usual sense (all maps with at least one argument in the ideal take values in the ideal). 
The quotient space (\ref{Kquotient}) then automatically inherits a BV$_{\infty}^\B$ algebra.

To this end we redefine  $\mu$ by a $C_\infty$ isomorphism: 
\begin{equation}\label{m2} 
m_2:=\mu+\partial f_2\;,    
\end{equation}
where $f_2$ is a graded symmetric bilinear map of degree $-1$, and we have introduced the  shorthand notation 
\begin{equation}
\partial f_2 :=Q\circ f_2+f_2\circ(Q\otimes 1+1\otimes Q)\;.   
\end{equation} 
In the realm of  $C_\infty$ algebras,  the shifted product (\ref{m2}) should be considered equivalent to $\mu$ 
for any $f_2$. There are several equivalent choices of $f_2$ that make $\cI$ an ideal with respect to (\ref{m2}). 
For instance, setting $f_2(F_1,F_2)=c\,\theta_2(F_1,F_2)$,  with $\theta_2$ a graded symmetric bilinear map 
of degree $G=-2$, we can take, with the $\chi$'s in $\cI$ and $F$ in its complement,  
\begin{equation}\label{theta2 glory}
\begin{split}
\theta_2(F,\chi) &=(-1)^F \frac{\del F}{\del C^\mu} \frac{\del\chi}{\del C_\mu}\;, \\
\theta_2(\chi_1,\chi_2)&=-\tfrac12\,\cM\,\frac{\del^2\chi_1}{\del C^\mu\del C^\nu}\cdot \frac{\del^2\chi_2}{\del C_\mu\del C_\nu}\;, 
\end{split}    
\end{equation}
while  $\theta_2$ is zero otherwise. 
Since $\cI$  is then an ideal with respect to the resulting $m_2$,  the quotient space $\cK$ 
inherits the $C_\infty$ product 
\begin{equation}
m_2\big([F_1],[F_2]\big):=[m_2(F_1,F_2)]\;.     
\end{equation}
The non-vanishing components of $m_2$ in $\cK$ are found to be 
\begin{equation}
\begin{split}
m_2(\lambda_1,\lambda_2)&=\lambda_1\lambda_2\;,\\
m_2(\lambda,\cA)&=C^\mu(\lambda A_\mu)+c(\lambda\varphi+A^\mu\del_\mu\lambda)\;,\\
m_2(\lambda,\cE)&=c\,C^\mu(\lambda E_\mu)+\cM(\lambda E)\,,\\
m_2(\lambda,N)&=c\,\cM(\lambda N)\,,\\
m_2(\cA_1,\cA_2)&=2c\,C_\mu
\big(\varphi^{}_{[1} A^\mu_{2]}+2\,A_{[1}^\rho\del_\rho A^\mu_{2]}+\del^\mu A_{[1}^\rho A^{}_{2]\rho}\big)\,,\\
m_2(\cA,\cE)&=c\,\cM(\varphi E-A^\mu E_\mu+2\,A^\mu\del_\mu E)\,,
\end{split}    
\end{equation}
where we denoted by $\cA=(A_\mu,\varphi)$ and $\cE=(E_\mu,E)$ the doublets of fields and their corresponding equations.

Due to the shift by $f_2$ in  (\ref{m2}), the resulting algebra $(\Omega^{\bullet},Q, m_2)$ is no longer a strict 
dgCom algebra. The associativity of $m_2$ is violated by $f_2$ contributions, but it is homotopy associative in 
the sense of $C_\infty$ algebras. The needed trilinear  3-product is given by 
\begin{equation}\label{m3}
\begin{split} 
m_3 &:=\mu\circ( f_2\otimes1-1\otimes  f_2)+ f_2\circ(\mu\otimes1-1\otimes \mu)\\
 & \quad\; +\del f_2\circ (f_2\otimes1-1\otimes  f_2)\;.    
\end{split} 
\end{equation}
Remarkably, $\cI$ is still an ideal with respect to this  $m_3$. As a consequence,  no further shifts and no higher 
$m_n$ with $n\geq 4$
are needed. The  space  $(\Omega^{\bullet},m_1:=Q, m_2,m_3)$ then defines a $C_{\infty}$ algebra, 
and $\cI$ is an ideal of this $C_{\infty}$ algebra. Consequently, the quotient space (\ref{Kquotient}) inherits 
a $C_{\infty}$ algebra with highest non-trivial product being $m_3$, whose only non-vanishing component in $\cK$ takes three vector fields $A_i^\mu$ as inputs and reads
\begin{equation}
m^\mu_3(\cA_1,\cA_2,\cA_3)=A_1^\rho A_{2\rho}A_3^\mu+A_3^\rho A_{2\rho}A_1^\mu-2\,A_1^\rho A_{3\rho}A_2^\mu.    
\end{equation}
This is the kinematic $C_{\infty}$ algebra of Yang-Mills theory \cite{Zeitlin:2008cc,Borsten:2021hua,Bonezzi:2022yuh} 
that, thereby, we have derived from a simple algebra of superforms.

\section{Kinematic algebra} 

As mentioned  in the introduction, the $C_{\infty}$ algebra on $\Omega^{\bullet}$, and hence 
the $C_{\infty}$ algebra on the quotient $\cK$, is just the first layer of the full BV$_{\infty}^\B$ algebra 
underlying CK duality, to which we will turn now. 
In addition to the graded symmetric and homotopy associative product $m_2$, the BV$_{\infty}^\B$ algebra 
features the `Lie-type' bracket: 
 \begin{equation}\label{kinematicb2}  
  b_2:= [b,m_2]\;, 
 \end{equation} 
which encodes the failure of $b$ to act as a derivation on $m_2$. For a strict  BV$^\B$ algebra, $b_2$ is the 
antibracket, which is a graded Lie bracket satisfying the graded Jacobi identity and a Poisson compatibility condition 
with $m_2$. However, since here 
$m_2$ is only associative up to homotopy and since $b$ commutes with $Q$ to $\B$, c.f.~(\ref{QbBox}), 
the Jacobi identity and Poisson relation get  corrected. This defines the BV$_{\infty}^\B$ algebra. 
Specifically, the Jacobiator of $b_2$, i.e., its failure to obey the Jacobi identity, obeys 
\begin{equation}\label{BVJacobiii}
{\rm Jac}_{b_2}+\del b_3+6\,\theta_3\circ(\del^\mu\otimes\del_\mu\otimes 1)\circ\pi=0\;, 
\end{equation}
where $b_3$ is the three-bracket of the kinematic algebra, $\theta_3$ is a new map related to relaxing the Poisson compatibility, and 
$\pi$ is a certain Young-tableaux projection \cite{Bonezzi:2022bse}. It is straightforward to determine $b_3$ and $\theta_3$ for the full BV$_{\infty}^\B$ algebra on $\Omega^{\bullet}$, but the challenge is to find a representative in  the class of maps related by  BV$_{\infty}^\B$ isomorphisms for which $\cI$ is an ideal. Once such a representative has been identified 
the quotient space $\cK ={\Omega^{\bullet}}/{\cI}$ in 
(\ref{Kquotient}) automatically 
inherits a BV$_{\infty}^\B$ algebra. 
We will comment in the conclusions on the ongoing research program to determine  
these maps algorithmically. 
   
For the remainder of this subsection we explain 
how the above relations 
are related to CK duality, for which 
the following formulation 
due to Mafra and Schlotterer is convenient \cite{Mafra:2014oia,Mafra:2015vca,Lee:2015upy, Garozzo:2018uzj,Bridges:2019siz,Mafra:2022wml}. One considers the so-called partial or color-ordered amplitudes, which  can be computed from 
color-ordered Feynman rules (or, equivalently, via  homotopy transfer of the kinematic $C_\infty$ algebra \cite{Nutzi:2018vkl,Arvanitakis:2019ald,Jurco:2019yfd,Arvanitakis:2020rrk,Bonezzi:2023xhn}). One then defines the 
Berends-Giele currents \cite{Berends:1987me} or multiparticle fields $\cA^\mu_{12\dots n}$, with  $i=1,2,\dots n$ single-particle labels, as  given by the same color-ordered Feynman diagrams, but with 
one leg kept off-shell and dressed with the corresponding propagator.
In terms of the local numerators of these diagrams (multiparticle polarizations) CK duality means that  the numerators obey the Lie symmetries of nested commutators, where the composition of commutators is dictated by the structure of the cubic diagram. For instance, the numerator $N^\mu_{12\dots n}$ associated with the cubic half-ladder diagram below should have the same symmetry as  $[[[T_1,T_2],\cdots],T_n]$ in its particle labels. 

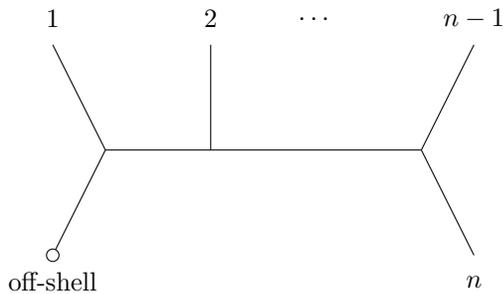
\begin{center}
    \begin{tikzpicture}[scale=0.7]
	\begin{pgfonlayer}{nodelayer}
		\node [style=off shell] (0) at (-4, -2) {};
		\node   (1) at (-3, 0) {};
		\node   (2) at (-1, 0) {};
		\node   (3) at (1, 0) {};
		\node   (4) at (3, 0) {};
		\node   (5) at (4, -2) {};
		\node   (6) at (-4, 2) {};
		\node   (7) at (-1, 2) {};
		\node   (8) at (1, 2) {};
		\node   (9) at (4, 2) {};
		\node   (10) at (-4, 2.5) {$1$};
		\node   (11) at (-1, 2.5) {$2$};
		\node   (12) at (1, 2.5) {$\dots$};
		\node   (13) at (4, 2.5) {$n-1$};
		\node   (14) at (4, -3) {};
		\node   (15) at (4, -2.5) {$n$};
		\node   (16) at (-4, -3) {};
		\node   (17) at (-4, -2.5) {off-shell};
	\end{pgfonlayer}
	\begin{pgfonlayer}{edgelayer}
		\draw (6.center) to (1.center);
		\draw (0) to (1.center);
		\draw (1.center) to (4.center);
		\draw (7.center) to (2.center);
		\draw (4.center) to (9.center);
		\draw (4.center) to (5.center);
	\end{pgfonlayer}
\end{tikzpicture}

\end{center}

To understand this in more detail, note from the 
action (\ref{cinftyAction})  that the color-ordered cubic and quartic vertices are described by the two- and three-products $m_2$ and $m_3$ of the $C_\infty$ algebra, while the propagator is given by $\frac{b}{\B}$.  Upon gauge fixing, the fields obey $b(A)=0$, so that the numerator contribution $b m_2(A_i,A_j)=b_2(A_i,A_j)$ coincides with the kinematic bracket (\ref{kinematicb2}). In particular, the numerator of a cubic half-ladder diagram is given by the successive nesting $b_2(b_2(\cdots)\cdots)$ of two-brackets. Thus, if $b_2$ were a strict Lie bracket, it would immediately follow that the numerator has the required Lie symmetries, hence implying CK duality.  

However, $b_2$ does  not define a strict Lie algebra, but its failure disappears at the amplitude level, at least  at four points. 
As an example, consider the color-ordered four-point amplitude  related to the three-particle field $\cA^\mu_{123}$, given by the following diagrams:
\begin{center}
    \begin{tikzpicture}[scale=0.25]
	\begin{pgfonlayer}{nodelayer}
		\node   (94) at (-4, 4) {};
		\node   (95) at (0, 0) {};
		\node   (96) at (4, 4) {};
		\node   (97) at (2, 4) {};
		\node   (98) at (-2, 4) {};
		\node   (99) at (-2, 2) {};
		\node   (100) at (2, 2) {};
		\node   (101) at (-3, 3) {};
		\node   (102) at (3, 3) {};
		\node [style=off shell] (103) at (0, -2) {};
		\node   (104) at (0, 4) {};
		\node   (105) at (-1, 3) {};
		\node   (106) at (1, 3) {};
		\node   (107) at (-4, 5) {};
		\node   (108) at (0, 5) {};
		\node   (109) at (4, 5) {3};
		\node   (110) at (-4, 5) {1};
		\node   (111) at (0, 5) {2};
		\node   (112) at (-1, 1) {};
		\node   (113) at (1, 1) {};
		\node   (114) at (-0.5, -1) {};
		\node   (115) at (0.5, -1) {};
		\node   (116) at (-1.5, 1.5) {};
		\node   (117) at (1.5, 1.5) {};
		\node   (118) at (-5, 0) {};
		\node   (119) at (-5, 0) {$\mathcal{A}^\mu_{123}=$};
		\node   (120) at (6, 4) {};
		\node   (121) at (10, 0) {};
		\node   (122) at (14, 4) {};
		\node   (123) at (12, 4) {};
		\node   (124) at (8, 4) {};
		\node   (125) at (12, 2) {};
		\node   (126) at (12, 2) {};
		\node   (127) at (7, 3) {};
		\node   (128) at (13, 3) {};
		\node [style=off shell] (129) at (10, -2) {};
		\node   (130) at (10, 4) {};
		\node   (131) at (9, 3) {};
		\node   (132) at (11, 3) {};
		\node   (133) at (6, 5) {};
		\node   (134) at (10, 5) {};
		\node   (135) at (14, 5) {3};
		\node   (136) at (6, 5) {1};
		\node   (137) at (10, 5) {2};
		\node   (138) at (9, 1) {};
		\node   (139) at (11, 1) {};
		\node   (140) at (9.5, -1) {};
		\node   (141) at (10.5, -1) {};
		\node   (142) at (8.5, 1.5) {};
		\node   (143) at (11.5, 1.5) {};
		\node   (144) at (5, 0) {};
		\node   (145) at (5, 0) {+};
		\node   (146) at (16, 4) {};
		\node   (147) at (20, 0) {};
		\node   (148) at (24, 4) {};
		\node   (149) at (22, 4) {};
		\node   (150) at (18, 4) {};
		\node   (151) at (20, 0) {};
		\node   (152) at (22, 2) {};
		\node   (153) at (17, 3) {};
		\node   (154) at (23, 3) {};
		\node [style=off shell] (155) at (20, -2) {};
		\node   (156) at (20, 4) {};
		\node   (157) at (19, 3) {};
		\node   (158) at (21, 3) {};
		\node   (159) at (16, 5) {};
		\node   (160) at (20, 5) {};
		\node   (161) at (24, 5) {3};
		\node   (162) at (16, 5) {1};
		\node   (163) at (20, 5) {2};
		\node   (164) at (19, 1) {};
		\node   (165) at (21, 1) {};
		\node   (166) at (19.5, -1) {};
		\node   (167) at (20.5, -1) {};
		\node   (168) at (18.5, 1.5) {};
		\node   (169) at (21.5, 1.5) {};
		\node   (170) at (15, 0) {};
		\node   (171) at (15, 0) {+};
		\node   (173) at (-2.5, 2.5) {};
		\node   (174) at (-3.5, 3.5) {};
	\end{pgfonlayer}
	\begin{pgfonlayer}{edgelayer}
		\draw (103) to (95.center);
		\draw (95.center) to (94.center);
		\draw (95.center) to (96.center);
		\draw (99.center) to (104.center);
		\draw (129) to (121.center);
		\draw (121.center) to (120.center);
		\draw (121.center) to (122.center);
		\draw (125.center) to (130.center);
		\draw (155) to (147.center);
		\draw (147.center) to (146.center);
		\draw (147.center) to (148.center);
		\draw (151.center) to (156.center);
	\end{pgfonlayer}
\end{tikzpicture}

\end{center}
which reads analytically 
\begin{equation}\label{3particleField} 
\begin{split}
\cA_{123}&= \frac{b}{s_{123}}\Big(\frac{m_2(bm_2(A_1,A_2),A_3)}{s_{12}}+\frac{m_2(A_1,bm_2(A_2,A_3))}{s_{23}}\\
&\qquad \qquad +m_3(A_1,A_2,A_3)\Big)\;,   \end{split}      
\end{equation}
where we have introduced the Mandelstam variables 
 \begin{equation}
     s_{i_1\dots i_n}=(k_{i_1}+\dots+k_{i_n})^2\;, 
 \end{equation}
 and all single particle fields $A^\mu_i$ are on-shell and transverse, with definite momentum $k_i^\mu$.
 One next uses that the Young-tableaux 
 symmetries of $m_3$ are such that one may write 
 \begin{equation}
m_3(A_1,A_2,A_3)=m_{3h}(A_1,A_2|A_3)-m_{3h}(A_2,A_3|A_1)
\;,    
\end{equation}
where $m_{3h}$ is antisymmetric in the first two inputs, in terms of which 
we have 
$[b,m_{3h}]=-\theta_{3h}$ \cite{Bonezzi:2022bse}.  
Introducing 
the three-particle polarizations 
\begin{equation}
N_{ijk}:=b_2(b_2(A_i,A_j),A_k)-s_{ij}\,\theta_{3h}(A_i,A_j|A_k)\,, 
\end{equation}
the three-particle field (\ref{3particleField}) is then given by 
\begin{equation}
\cA_{123}=\frac{1}{s_{123}}\,\left(\frac{N_{123}}{s_{12}}-\frac{N_{231}}{s_{23}}\right)\;.     
\end{equation}
CK duality now requires $N_{ijk}=-N_{jik}$, which holds by definition, and $N_{123}+N_{231}+N_{312}=0$. This sum is given by the Jacobiator of $b_2$, plus the terms with $\theta_3$. In BV$_\infty^\B$, the two-bracket obeys the  deformed homotopy Jacobi identity (\ref{BVJacobiii}), and using the fact that $s_{12}+s_{23}+s_{31}=s_{123}$, the sum of the numerators becomes
\begin{equation}
N_{123}+N_{231}+N_{312}=-\del b_3(A_1,A_2,A_3) \;. 
\end{equation}
The term on the right-hand side is exact and drops out from the amplitude, 
thereby establishing CK duality at the level of four-point amplitudes. In the framework of \cite{Mafra:2015vca,Lee:2015upy}, the exact term $\del b_3(A_1,A_2,A_3)$ is viewed as a multiparticle gauge transformation, used to shift the numerators as $N_{123}\rightarrow N_{123}+\frac13\,\del b_3(A_1,A_2,A_3)$ in order to obey CK duality with one off-shell leg.\\

\section{Summary \& Outlook}

In this letter we have introduced a new paradigm for deriving the `kinematic algebra' of Yang-Mills theory, starting  from the  algebra of differential forms on a superspace. While it has been appreciated in recent years that
the algebraic structure underlying color-kinematics duality is a homotopy generalization of a Batalin-Vilkovisky (BV) algebra, named BV$_\infty^\B$ by Reiterer \cite{Reiterer:2019dys}, so far it has remained elusive to define this algebra for standard Yang-Mills theory explicitly to all orders. Ideally one would like to obtain this algebra by a derived construction from a strict algebra that carries only linear and bilinear maps. 

Here we have identified an algebra of differential forms on a superspace that carries a strict BV$^\B$ algebra. 
Remarkably, this space of superforms  contains the kinematic space  of genuine Yang-Mills theory as a subspace. 
Moreover, its complement $\cI$ within 
the full space is an `ideal up to homotopy'. 
The $C_{\infty}$ algebra of Yang-Mills theory is then realized as the quotient algebra. 

It remains to generalize this derived construction of  the $C_{\infty}$ subsector to the full BV$_\infty^\B$ algebra. One challenge here is to realize the BV$_\infty^\B$ isomorphism 
on the full space that transforms all higher maps to a form where $\cI$ is an ideal. 
To this end the recent work \cite{ValletteNew} in 
pure mathematics might be important. 
Establishing an effective procedure for obtaining these transformations would yield an algorithm for determining the CK duality obeying kinematic numerators in a manifestly local fashion, thereby proving CK duality of Yang-Mills theory at tree-level. It would also be interesting to generalize this to more complicated Yang-Mills theories such as supersymmetric theories \cite{Bonezzi:2025anl,Bonezzi:2025bvl}. 
 
The algorithmic determination of the higher BV$_\infty^\B$  maps would also be crucial in order to realize gravity as the double copy of Yang-Mills theory. 
Specifically, one would like to obtain the off-shell and local Lagrangian of gravity, and its diffeomorphism transformations, purely from the Yang-Mills data.
(See  \cite{Anastasiou:2014qba,Anastasiou:2018rdx,Borsten:2020xbt,Borsten:2020zgj,Diaz-Jaramillo:2021wtl} for earlier work.) 
A promising short-cut might be to double copy the full algebra of superforms as in \cite{Borsten:2023ned,Bonezzi:2024dlv,Ben-Shahar:2025dci} and then to determine ${\cal N}=0$ supergravity,  respectively double field theory \cite{Siegel:1993th,Hull:2009mi,Hohm:2010jy,Hohm:2010pp} as in \cite{Diaz-Jaramillo:2021wtl}, 
as the appropriate quotient.

\subsection*{Acknowledgements}

We would like to thank Giuseppe Casale, Felipe D\'iaz-Jaramillo, Jan Pulmann, Michael Reiterer and Bruno Vallette for stimulating discussions, correspondence and collaboration in related projects.
This work is supported by the Deutsche Forschungsgemeinschaft (DFG, German Research Foundation) - Projektnummer 417533893/GRK2575 ``Rethinking Quantum Field Theory" and Projektnummer 9710005691, ”Homological Quantum Field Theory”.

\end{document}